\documentclass[12pt]{article}
\usepackage[dvips]{color}
\usepackage{epsfig}
\usepackage{amsmath}
\usepackage{cite}
\usepackage{graphicx}
\usepackage{color}
\usepackage{subfigure}

\begin{document}
\begin{center}
\Large{\bf Weak Gravity Conjecture of Charged-Rotating-AdS Black Hole Surrounded by Quintessence and String Cloud}\\
\small \vspace{1cm} {\bf J. Sadeghi$^{a,e}$\footnote {Email:~~~pouriya@ipm.ir}},
{\bf S. Noori Gashti$^{a}$\footnote {Email:~~~saeed.noorigashti@stu.umz.ac.ir}},
{\bf \.{I}. Sakall{\i}$^{b}$\footnote {Email:~~~izzet.sakalli@emu.edu.tr}},
{\bf B. Pourhassan$^{c,d,e}$\footnote {Email:~~~b.pourhassan@du.ac.ir}}
\\
\vspace{0.5cm}$^{a}${Department of Physics, Faculty of Basic
Sciences,\\ University of Mazandaran P. O. Box 47416-95447, Babolsar, Iran}\\
\vspace{0.5cm}$^{b}${Physics Department, Eastern Mediterranean
University, Famagusta 99628, North Cyprus via Mersin 10, Turkey.}\\
\vspace{0.5cm}$^{c}${School of Physics, Damghan University, Damghan, 3671641167, Iran.}\\
\vspace{0.5cm}$^{d}${Physics Department, Istanbul Technical University, Istanbul 34469, Turkey.}\\
\vspace{0.5cm}$^{e}${Canadian Quantum Research Center 204-3002 32 Ave Vernon, BC V1T 2L7 Canada.}\\
\small \vspace{1cm}
\end{center}
\begin{abstract}
A series of corrections to general relativity have recently been applied to find the relationship between entropy and extremality-bound black holes. This relationship has been investigated for many black holes, such as charged AdS, rotating, and massive gravity black holes. We give a minor constant correction to the action and confirm these universal relations for a charged-rotating-AdS black hole. We then examine these calculations for the black hole, surrounded by the quintessence and the cloud of string. In this paper,  we evaluate a new universal relation. It means that we find the relation between the extremal mass of the black hole and the factor of cloud string and observe that the corresponding universal relation is well established.
We find that the quintessence terms play a very effective role in calculating the mass-charge ratio and concept of weak gravity conjecture of black holes. We note here that the added constant correction is inversely related to the entropy of the black hole. It leads us to see that the mass-to-charge ratio decreases and fully confirms the black hole's weak gravity conjecture (WGC).
\end{abstract}
\newpage
\section{Introduction}
Black holes are the best objects for developing gravity as a form of quantum gravity. Recently, quantum gravity has been studied from different perspectives, such as the framework of low-energy effective field theories. In that case, string theory provides a complete description of quantum gravity. Therefore, to study of quantum gravity in string theory, various conjectures and methods are always used, including the swampland program and weak gravity conjecture \cite{1,2,3,4,5,6,7,8,9,10,11,12,13,14,15,16,17,18,19, JHAP}. If we want to have quantum gravity as a form of low-energy theories, we need some object as a form of black hole background with special constraints as the charge-to-mass ratio must always be greater than one $Q\geq M$. This will be the best constraint for the extremal black holes to evaporate. If this conjecture is incorrect, the concepts related to the black hole evaporation and many other concepts that researchers have approved will face major problems. For example, one of these issues is the violation of cosmic censorship. The development of this conjecture can suggested by the string theory for introducing new way to find some compatible theory \cite{20,21,22,23,24,26,27,28,29,30,31,32,33,34}. As mentioned in the text, many attempts have been made to calculate universal relations in recent years. Recently Goon and Penco \cite{35} have presented a universal thermodynamic relation due to the perturbative corrections to the thermodynamic relations. This universal relation has been proven for charged AdS black holes and investigated in other works related to rotating as well as massive gravity black holes \cite{36,37,38,39}. We want to prove this universal relation for a Kerr-Newman-AdS black hole surrounded by quintessence and the cloud of the string regarding all the motivations mentioned earlier. After that, we will try to compare the universal relation of different forms of black holes.\\
Before going further, we first give some review of these physical subjects. As we know, all current observations represent an expanding universe that has acceleration due to negative pressure. This negative pressure can have interpretations such that one of these interpretations is quintessence \cite{40,41,42,43,44, Q}. Quintessence is described by a specific scalar field, which is minimally coupled with gravity due to the anti-gravity nature of dark energy, which always has specific potentials that lead to late time inflation. In the case of extremal black holes surrounded by quintessence, particular changes can be considered, such as black holes that do not have a singularity \cite{44,45,46}.\\
Another case we consider here is the black hole with the cloud of string, and the analysis of these string clouds was first reviewed by Letelier \cite{47}. He studied Schwarzschild's expansion kind of black hole is investigated by \cite{47}. Also, the solutions of a black hole surrounded by a spherical symmetry cloud of the string are studied by \cite{48,49}.\\
In this article, our primary goal is to explore a new implication for this string cloud. We are looking for a new universal thermodynamic relation for the corresponding system. All of the above concepts motivate us to confirm universal relations for the charged-rotating-AdS black hole. We first examine the universal relations for this black hole, so we add a small constant correction to the action and obtain the modified thermodynamic relations. By analyzing the thermodynamics relations, we get the universal relation. Then, considering the quintessence, we calculate all the mentioned steps for this black hole. Then, we consider the cloud of string as a new feature for Kerr-Newman-AdS black hole. We investigate all universal relations; primarily, we evaluate a new universal relation. It is the relation between the mass of the black hole and the factor related to the cloud of string, i.e., $$-\zeta\frac{\partial b}{\partial \epsilon}=\frac{\partial M_{ext}}{\partial\epsilon}$$
where $b$ is string cloud parameter and $\zeta$ assumes its conjugate in thermodynamic relation. By solving some complicated equations, we obtain the modified thermodynamic parameters like mass, entropy, etc. Here, we note that the black hole's entropy increases when the added correction constant has a negative value, the mass of the black hole and the mass-charge ratio decrease. Hence, these black holes show WGC-like behavior. Finally, when we compare the obtained results from the different black holes from a universal relation point of view, we will see that the additional correction constant plays a very influential role in the black hole's thermodynamic parameters.\\
All the above information motivates us to organize this article in this way. In section 2, we confirm the universal relations for the Kerr-Newman-AdS black hole due to a small constant correction. This calculation helps us obtain the new universal relation of an above-mentioned black hole surrounded by quintessence and quintessence with a cloud of string, discussed in sections 3 and 4, respectively. In section 5, we describe the results of universal relation and compare the different black holes' results. Finally, in section 6, we conclude and summarize the results.\\
{\section{The motivation, WGC, and universal relation}
The assumption of weak gravity conjecture is a tempting concept that has recently attracted much attention in various concepts of particle physics and cosmology.
This conjecture states that any theory of quantum mechanics consistent with Einstein-Maxwell must have superextremal particles.
It can be well extended to theories with the concept of multiple U(1) gauge fields \cite{4}. Of course, this generalization can also manifest itself in various other types of structures, such as p-form fields in various dimensions and holographic setups \cite{b}.\\
One of the most important features or the great importance of this conjecture, and perhaps, in other words, one of the most important benefits that this concept has in itself, is the introduction of a series of specific criteria for when dealing with low-energy models that be banished to the swampland \cite{c}.
Thus, these criteria provide the conditions under which we can limit our attention to models which potentially have a consistent UV embedding \cite{2,4,b}.
These constraints, which apply to models with multiple axions, are in the form of a correlated boundary on the axion field range and instanton action, which leads to the creation of axion potential.
The weak gravity conjecture also imposes some limitations on monodromy models \cite{d}.
Other recent studies based on this statement include the study of non-supersymmetric vacuum stability \cite{2}
Of course, this concept is still being studied in depth to clarify a deeper understanding of this statement.
One of the most important motivations that led to the introduction and study of weak gravity conjecture is avoiding remnants and the closely related species problem \cite{f}, which you can see in \cite{2,3,h, i, j,l}for further study.\\
The weak gravity conjecture states that in self-consistent theories of quantum gravity, the strength of gravity is limited from above by various gauge forces.
This statement claims to be able to describe theories in which a U(1) gauge field is coupled consistently to gravity.
Thus, there must be a particle whose proper mass is limited by its charge according to the Planck unit, i.e., $ M/ M_{P} <q $ \cite{2}.
This attractive conjecture has recently lured the attention of many physicists, and a lot of work has been done in this field and in connection with different types of black holes \cite{38}.
But the valuable point to note here is that despite the variety of field theories and string theories that support this statement, it is like a toddler that does not yet have much evidence for its validity.
 Of course, a lot of efforts have been made to validate it, including Shahar Hod in \cite{k} proving that this statement can be deduced from Bekenstein generalized second law of thermodynamics.
Recently, this statement has been used to study the structure of extremal black holes, which has yielded exciting results.\\
Of course, in the following article, we have somehow challenged the relationship between the black hole, the universal relationship, and the weak gravity conjecture.
Universal relations are considered an essential concept in various theories, which is perhaps one of the biggest concerns of the scientific community, the unity and integration of the fundamental forces of nature or achieving a global relationship.
Recently, universal relations in various physical concepts have been considered, and efforts have been made,
including the universal relationship presented by Goon-Penco \cite{35} in connection with recent thermodynamic relationships. With respect to this fact that a series of perturbative corrections to general relativity leads to changes in the entropy of black holes and their extremality boundaries. Thus, they proved a series of universal relationships between leading modifications to these quantities through exciting calculations.
They proved that when entropy corrections are assumed to be positive, the result leads to a perturbation decrease in the mass of the extremal black holes, provided one takes other extensive variables to be constant.
Thus, considering perturbative corrections about black holes, they concluded that extremality relations of a wide range of black holes could exhibit weak gravity conjecture-like behavior.\\
As mentioned earlier, a thermodynamic extremality relation has recently been introduced in \cite{35}, which is expressed in the following form.
\begin{equation*}\label{1}
\frac{\partial M_{ext}(\overrightarrow{\mathcal{Q}}, \varepsilon)}{\partial\varepsilon}=\lim_{M\rightarrow M_{ext}}-T\bigg(\frac{\partial S(M, \overrightarrow{\mathcal{Q}}, \varepsilon)}{\partial\varepsilon}\bigg)_{M, \overrightarrow{\mathcal{Q}}},
\end{equation*}
where $\varepsilon$, $M_{ext}$, $T$, and $S$ are the control parameters of the correction, mass bound, temperature, and entropy of the black hole after the correction, respectively.
Also, $\overrightarrow{\mathcal{Q}}$ is introduced as extensive quantities of the black hole thermodynamics.
Details of the concepts and extensive mathematics associated with these concepts can be found in \cite{35}.
Hence all of the above concepts became the impetus for this study to be a special kind of tiny perturbations that are somehow added to the part AdS of black holes in action. Due to changes in entropy values, and the extremal black hole boundary, we are obtained to create a new type of this universal relation about other thermodynamic values, and the connection of all these values to each other.
 It can also be a valuable point that the presence of perturbative corrections, however small, can lead to a series of universal relationships between different thermodynamic values.
 Of course, this can be a valuable point for further consideration of other concepts with regard to Perturbative corrections.}

\section{Kerr-Newman-AdS black hole}
This section will investigate the universal relation and show the weak gravity conjecture of how to come to the game
in the corresponding black holes. To study the universal relation,  we have to consider generally thermodynamic relations
such as temperature and mass, and angular velocity according to the solution of action, which is given by the first law of black hole thermodynamics.\\
On the other hand, we need some black hole solutions to prove the universal relations from thermodynamics quantities. So, for these reasons, we introduce the Einstein-Maxwell-AdS action in four dimensions, which  is given by \cite{50,51},
\begin{equation}\label{1}
\mathcal{I}=-\frac{1}{16\pi G}\int_{M}d^{4}x\sqrt{-g}(R-F^{2}+2\Lambda),
\end{equation}
{where $G$ is the Newton's Gravitational Constant}, $F=dA$ is field strength, $A$ is the potential 1-form and $\Lambda=-\frac{3}{l^{2}}$ is cosmological constant {with $l$ the radius of AdS space}. The solution of action (\ref{1}) will be Kerr-Newman AdS black hole which is given by \cite{50,51},
\begin{eqnarray}\label{2}
ds^{2}&=&-\frac{f(r)}{\rho^{2}}\left(dt^{2}-\frac{a\sin^{2}\theta}{\Xi}d\phi\right)^{2}+\frac{\rho^{2}}{f(r)}dr^{2}\nonumber\\
&+&\frac{\rho^{2}}{g(\theta)}d\theta^{2}+\frac{g(\theta)\sin^{2}\theta}{\rho^{2}}\left(a dt-\frac{r^{2}+a^{2}}{\Xi}d\phi\right)^{2},
\end{eqnarray}
where the metric functions defined as,
\begin{eqnarray}\label{3}
f(r)&=&r^{2}-2Mr+a^{2}+Q^{2}+\frac{r^{2}}{l^{2}}(r^{2}+a^{2}),\nonumber\\
g(\theta)&=&1-\frac{a^{2}}{l^{2}}\cos^{2}\theta,
\end{eqnarray}
with
\begin{eqnarray}\label{3.1}
\Xi&=&1-\frac{a^{2}}{l^{2}}\nonumber\\
\rho^{2}&=&r^{2}+a^{2}\sin^{2}\theta,
\end{eqnarray}
Where $M$,  $Q$, and $a$ are the black hole's mass, charge, and rotational parameters. The outer and inner event horizons associated with a black hole are calculated from $f(r)=0$.

\begin{figure}[h!]
\begin{center}$
\begin{array}{cccc}
\includegraphics[width=55 mm]{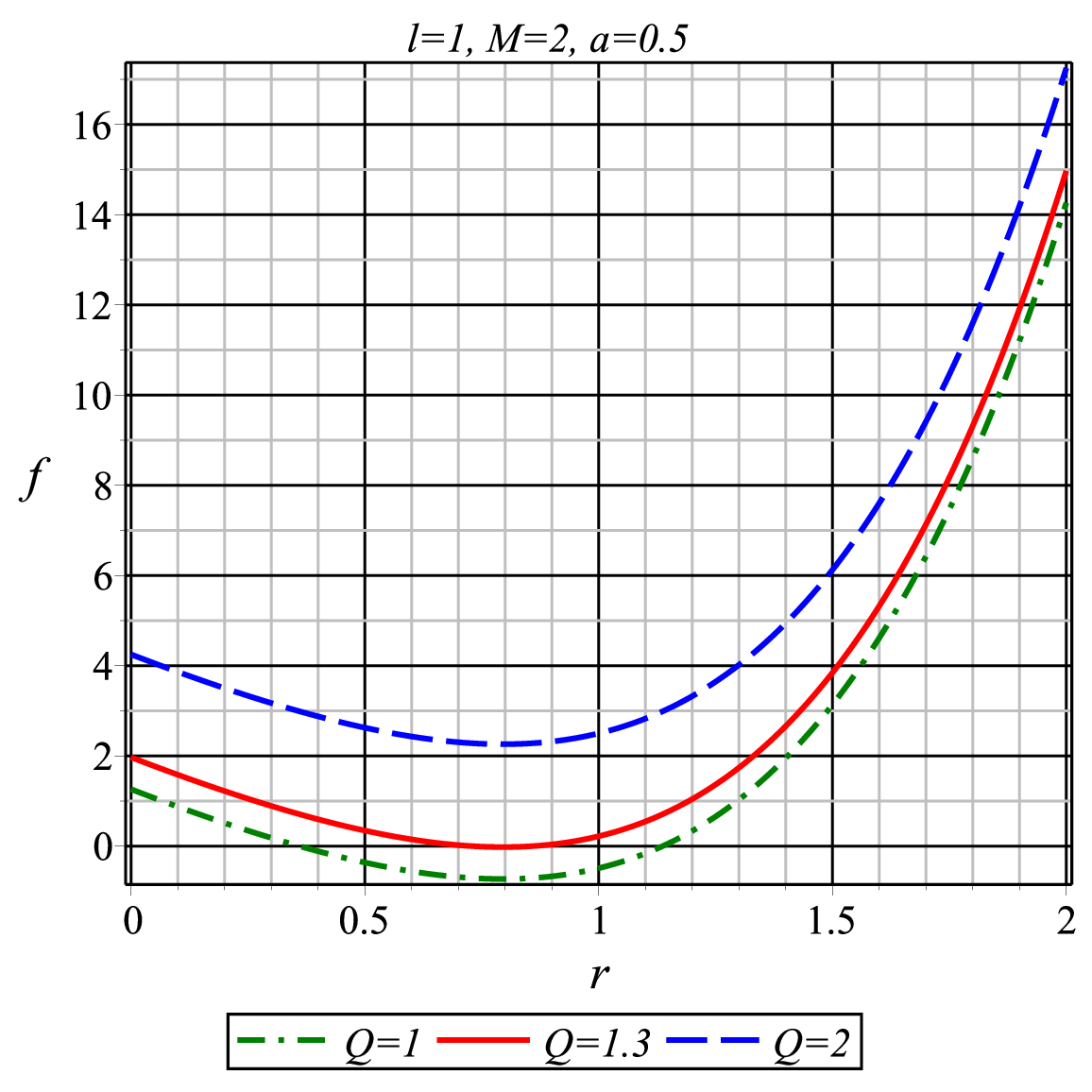}
\end{array}$
\end{center}
\caption{Horizon structure of Kerr-Newman AdS black hole.}
\label{figf}
\end{figure}
In Fig. \ref{figf}, we can see three possibilities. There is a black hole with two event horizons $r_{\pm}$ (see dash-dotted green line of Fig. \ref{figf}). In the case of $r_{+}=r_{-}$, we have an extremal black hole with zero temperature (see the solid red line of Fig. \ref{figf}). Also, the naked singularity is possible for a large enough black hole charge (see dashed blue line of Fig. \ref{figf}).\\
{As the black hole entropy given by $ S=\frac{\pi l^{2}(r_{+}^{2}+a^{2})}{l^{2}-a^{2}}$, Now, to have functions in terms of entropy or $r_{+}\rightarrow S$, the result be an imaginary. So we will consider a solution for forthcoming our investigation,i.e.,  we assume the event horizon $r_{+}$ and the radius of AdS $l$ larger than the angular momentum per unit mass. So,  in this route, we will perform the calculations for studying the thermodynamic relation of Kerr-Newman-AdS block holes such as mass, temperature, and angular velocity, which are given by},
\begin{equation}\label{4}
M=\frac{a^{2}\sqrt{\pi}}{2\sqrt{S}}+\frac{\sqrt{\pi}Q^{2}}{2\sqrt{S}}+\frac{\sqrt{S}}{2\sqrt{\pi}}+\frac{a^{2}
\sqrt{S}}{2l^{2}\sqrt{\pi}}+\frac{S^\frac{3}{2}}{2l^{2}\pi^{\frac{3}{2}}},
\end{equation}
\begin{equation}\label{5}
T=-\frac{a^{2}\sqrt{\pi}}{4S^{\frac{3}{2}}}-\frac{\sqrt{\pi}Q^{2}}{4S^{\frac{3}{2}}}+\frac{1}
{4\sqrt{\pi}\sqrt{S}}+\frac{a^{2}}{4l^{2}\sqrt{S}\sqrt{\pi}}+\frac{3\sqrt{S}}{4l^{2}\pi^{\frac{3}{2}}},
\end{equation}
and
\begin{equation}\label{6}
\Omega=\frac{a\sqrt{\pi}}{\sqrt{S}}+\frac{a\sqrt{S}}{l^{2}\sqrt{\pi}}
\end{equation}
Now, we are going to give some small corrections as $\epsilon$ to the action (\ref{1}), so the modified form of the action will be as,
\begin{equation}\label{7}
\mathcal{I}=-\frac{1}{16\pi G}\int_{M}d^{4}x\sqrt{-g}(R-F^{2}+(1+\epsilon)\times2\Lambda).
\end{equation}
Due to the modification of the action, the black hole solution is also modified. Therefore, some thermodynamic quantities of black holes will also be modified. Hence, the modified mass and temperature are obtained by the following equations,
\begin{equation}\label{8}
M=\frac{a^{2}\sqrt{\pi}}{2\sqrt{S}}+\frac{\sqrt{\pi}Q^{2}}{2\sqrt{S}}+\frac{\sqrt{S}}{2\sqrt{\pi}}+
\frac{a^{2}\sqrt{S}}{2l^{2}\sqrt{\pi}}+\frac{(1+\epsilon)S^\frac{3}{2}}{2l^{2}\pi^{\frac{3}{2}}},
\end{equation}
\begin{equation}\label{9}
T=-\frac{a^{2}\sqrt{\pi}}{4S^{\frac{3}{2}}}-\frac{\sqrt{\pi}Q^{2}}{4S^{\frac{3}{2}}}+\frac{1}{4\sqrt{\pi}
\sqrt{S}}+\frac{a^{2}}{4l^{2}\sqrt{S}\sqrt{\pi}}+\frac{3\sqrt{S}(1+\epsilon)}{4l^{2}\pi^{\frac{3}{2}}} ,
\end{equation}
While angular velocity remains unchanged and given by the equation (\ref{6}).\\
According to the above-modified expressions, the mass and the temperature of a black hole are modified with a minor constant correction. It can be stated that when the added correction is continuously negative, the mass of the black hole decreases and the mass-charge ratio of the black hole decreases and approaches one. These changes of mass and the mass-charge ratio of the black hole are essentially a confirmation of the WGC. To obtain new universal relation, we use equation (\ref{8}) and write the correction parameter as,
\begin{equation}\label{11}
\epsilon=\frac{-a^{2}l^{2}\pi^{2}-l^{2}Q^{2}\pi^{2}+2l^{2}M\pi^{\frac{3}{2}}\sqrt{S}-a^{2}\pi S-l^{2}\pi S-S^{2}}{S^{2}}.
\end{equation}
Now, we take the derivative concerning $S$ according to the equation (\ref{11}), and obtain,
\begin{equation}\label{12}
\frac{\partial \epsilon}{\partial S}=\frac{\pi(a^{2}(2l^{2}\pi+S)+l^{2}(2\pi Q^{2}-3M\sqrt{\pi}\sqrt{S}+S))}{S^{3}}.
\end{equation}
Then, we use the equations (\ref{9}) and (\ref{12}) to obtain as expression for $T\frac{\partial S}{\partial \epsilon}$. In that case, we arrive long term; this leads us to have manipulation with some limitations for the related terms. All the above information helps us to simplify the obtained results. So, finally, one can obtain the following relation,
\begin{equation}\label{13}
-T\frac{\partial S}{\partial \epsilon}=\frac{S^{\frac{3}{2}}}{2l^{2}\pi^{\frac{3}{2}}}.
\end{equation}
To obtain the second part of the universal relation, we use the equations (\ref{8}) and (\ref{9}). In that case, the extremal mass helps us obtain the following equation,
\begin{equation}\label{14}
\frac{\partial M_{ext}}{\partial\epsilon}= \frac{S^{\frac{3}{2}}}{2l^{2}\pi^{\frac{3}{2}}}.
\end{equation}
We can see that, two equations (\ref{13}) and (\ref{14}) are the same. We first proved the Goon-Penco universal extremality relation for this black hole. Now, we are going to examine another universal relation. In that case, just like the previous results, we use the equation (\ref{11}). So,  we will have,
\begin{equation}\label{15}
\frac{\partial \epsilon}{\partial Q}=-\frac{2l^{2}\pi^{2}Q}{S^{2}}.
\end{equation}
By considering the electric potential as $\Phi=\frac{\sqrt{\pi}Q}{\sqrt{S}}$ from the Kerr-Newman-AdS black hole and assuming extermality bound, one can rewrite equation (\ref{15}) as,
\begin{equation}\label{16}
-\Phi\frac{\partial Q}{\partial \epsilon}=\frac{S^{\frac{3}{2}}}{2l^{2}\pi^{\frac{3}{2}}}.
\end{equation}
As we can see, the equations (\ref{16}) and (\ref{14}) are the same. In this way, another universal relation is also approved.\\
In the following, we will try to seek another universal relation. This universal relation coming from relation between mass and pressure $P=\frac{3}{8\pi l^{2}}=-\frac{\Lambda}{8\pi}$. Also, here we use  equation (\ref{11}) and write the following derivative,
\begin{equation}\label{17}
\frac{\partial \epsilon}{\partial P}=\frac{3(\pi(a^{2}+Q^{2})-2M\sqrt{\pi S}+S}{8P^{2}S^{2}}.
\end{equation}
Therefore, according to the thermodynamic relation related to the black hole volume such as $V=\frac{4}{3}a^{2}\sqrt{\pi S}+\frac{4(1+\epsilon)S^{\frac{3}{2}}}{3\sqrt{\pi}}$ as well as extremal bound, we will arrive the following equation,
\begin{equation}\label{18}
-V\frac{\partial P}{\partial \epsilon}=\frac{S^{\frac{3}{2}}}{2l^{2}\pi^{\frac{3}{2}}}.
\end{equation}
We also see that the two equations (\ref{18}) and (\ref{14}) are precisely the same. So, we see here that another universal relation is also confirmed.\\
Now, we are going to consider the rotation of the black hole and prove the last universal relation. In that case, we use the equation (\ref{11}) and allow the parameter of rotation to play an essential role in universal relation, so we have the following relation,
\begin{equation}\label{19}
\frac{\partial \epsilon}{\partial a}=\frac{-2al^{2}\pi^{2}-2a\pi S}{S^{2}}.
\end{equation}
To achieve the last universal relation, we take equation (\ref{6}) and (\ref{19}) for the corresponding black hole so that one can obtain,
\begin{equation}\label{20}
-\Omega\frac{\partial a}{\partial \epsilon}=\frac{S^{\frac{3}{2}}}{2l^{2}\pi^{\frac{3}{2}}}.
\end{equation}
Here, we see that two equations (\ref{20}) and (\ref{14}) are precisely the same.  We note here in the next section we obtain the universal relations for this black hole, which is surrounded by quintessence, and compare it with the results obtained in this section.\\

\begin{figure}[h!]
\begin{center}
\subfigure[]{
\includegraphics[height=6cm,width=6cm]{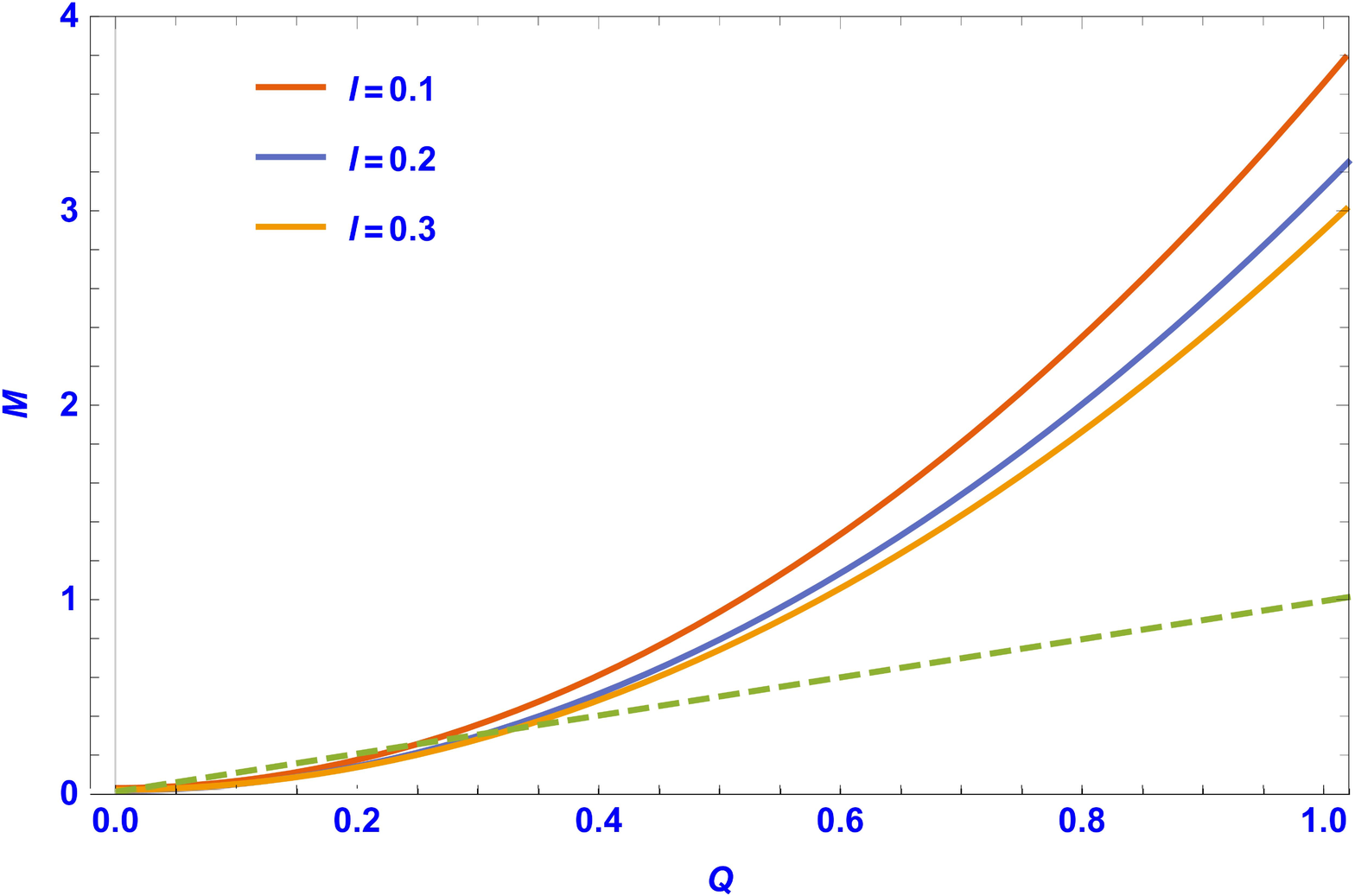}
\label{1a}}
\subfigure[]{
\includegraphics[height=6cm,width=6cm]{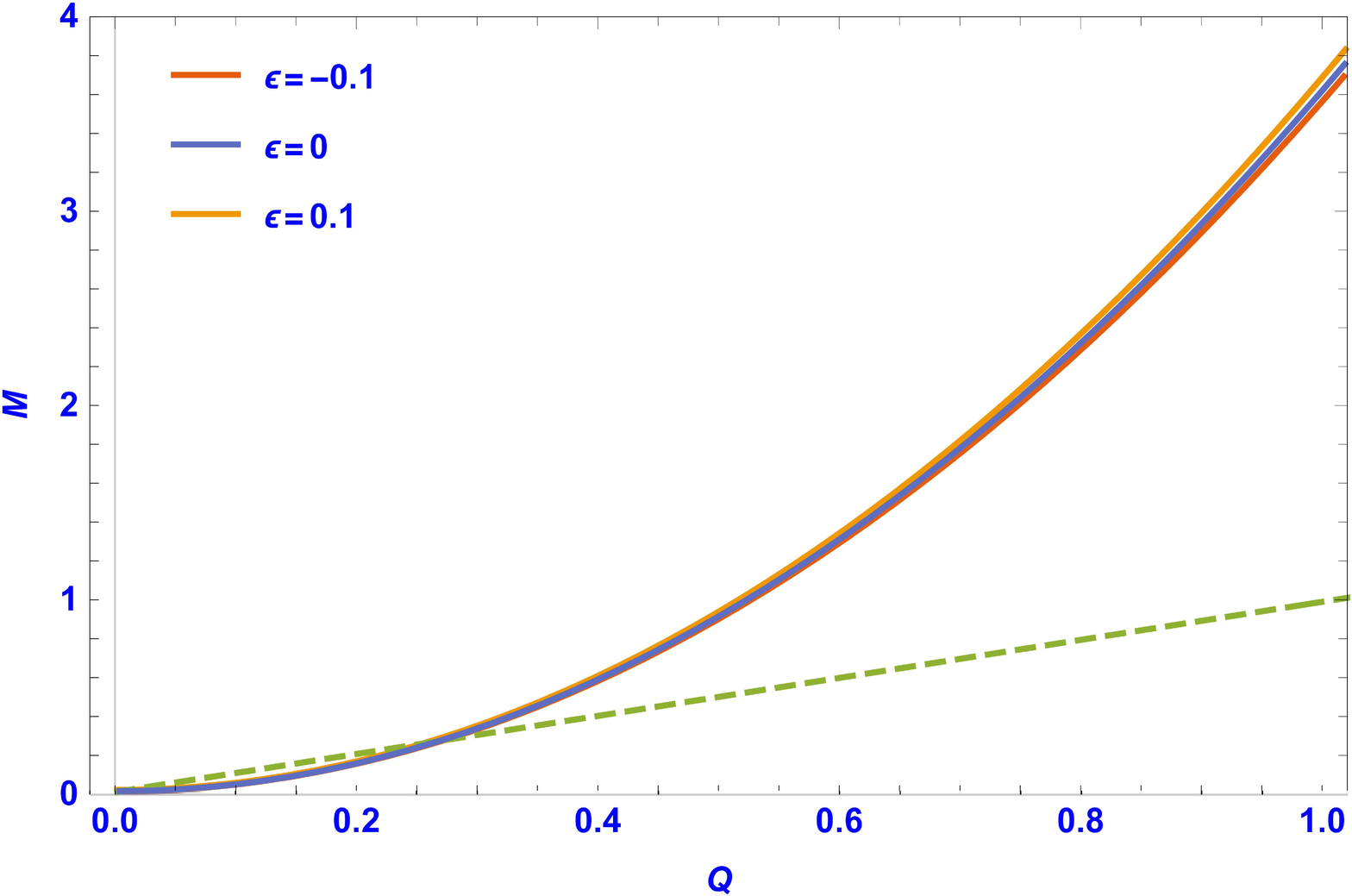}
\label{1b}}
\caption{\small{The plot of $M$ in terms of $Q$ for (a) $\epsilon=0$ (unmodified mass); (b) $l=0.1$ (modified mass). Dashed lines represent extremal case of Kerr-Newman-AdS black hole.}}
\label{1}
\end{center}
\end{figure}

The corresponding lines of Fig. \ref{1} lead us to compare two cases as unmodified and modified mass. Here, we fix some parameters and plot the mass in terms of charge $Q$ of black holes. The initial critical state is when the mass-to-charge ratio is one, shown as dashed lines of Fig. \ref{1}. It can be seen that the mass ratio of the unmodified black hole to the amount of charge is more than one. We consider different modes of AdS space radius $l= 0.1 ,0.2, 0.3$. As we can see in Fig. \ref{1} (a), the mass lines are different from the charge of the black hole for different modes. As we can see in Fig. \ref{1} (b), we compare the unmodified state by using a small constant correction for the mass. As shown in Fig. \ref{1} (b), when the constant correction is positive, the mass of the black hole increases, and when this constant correction is negative, the mass of the black hole decreases. Of course, this happens in reverse in the numerical calculation of entropy. It means that the entropy decreases with a positive correction. In fact, according to the concepts mentioned above, when we consider the small negative correction parameter, the mass of a black hole decreases and yields one, and the charge-to-mass ratio increases or the mass-charge ratio decreases, which is completely satisfied by the WGC.

\section{Kerr-Newman-AdS black hole with quintessence}
The previous section examined the universal relationships for the Kerr-Newman-AdS black hole without any additional terms. This section will confirm these universal relations for this black hole while surrounded by quintessence. So, the action for Kerr-Newman-AdS black hole surrounded by quintessence dark energy likewise the previous section is expressed in the following form \cite{51, 52}.
\begin{equation}\label{21}
\mathcal{I}=-\frac{1}{16\pi G}\int_{M}d^{4}x\{\sqrt{-g}(R-F^{2}+2\Lambda)+L_{q}\},
\end{equation}
where the $L_{q}$ is the Lagrangian of quintessence as a barotropic perfect fluid, which is given by \cite{53},
\begin{equation}\label{22}
L_{q}=-\rho\left(c^{2}+\int\frac{P(\rho)}{\rho^{2}}d\rho\right)=-\rho_{q}\left(1+\omega_{q}\ln(\frac{\rho_{q}}{\rho_{0}})\right),
\end{equation}
Where $\rho_{0}$ and $\rho_{q}$ are the constant parameter of integral and energy density, $ \omega_{q}$ is the barotropic index. The equation of state is given by $\frac{p_{q}}{\rho_{q}}=\omega_{q}$, which is bounded by $-1<\omega_{q}<-\frac{1}{3}$ for the quintessence dark energy. Of course,  the state equation is also bounded by  $-1>\omega_{q}$ for the phantom dark energy \cite{EPJP, CJP}. Therefore, the metric of Kerr-Newman-AdS black hole surrounded by quintessence dark energy is given by \cite{51,52},
\begin{equation}\label{23}
ds^{2}=\frac{\Sigma^{2}}{f(r)}dr^{2}+\frac{\Sigma^{2}}{g(\theta)}d\theta^{2}+\frac{g(\theta)\sin^{2}
\theta}{\Sigma^{2}}(a\frac{dt}{\Xi}-(r^{2}+a^{2})\frac{d\phi}{\Xi})^{2}-\frac{f(r)}{\Sigma^{2}}(\frac{dt}
{\Xi}-a\sin^{2}\frac{d\phi}{\Xi})^{2},
\end{equation}
where $f(r)$ is defined as,
\begin{eqnarray}\label{24}
f(r)&=&r^{2}-2Mr+a^{2}+Q^{2}-\frac{\Lambda}{3}r^{2}(r^{2}+a^{2})-\alpha r^{1-3\omega}\nonumber\\
g(\theta)&=&1+\frac{\Lambda}{3}a^{2}\cos^{2}\theta,
\end{eqnarray}
with
\begin{eqnarray}\label{24-2}
\Xi&=&1+\frac{\Lambda}{3}a^{2},\nonumber\\
\Sigma&=&r^{2}+a^{2}\cos^{2}\theta,
\end{eqnarray}
{where $\alpha$ is the quintessence parameter and $\omega$ is the quintessential state parameter}. Here, also one can write the thermodynamic quantities of Kerr-Newman-AdS block hole surrounded by quintessence such as mass, temperature, and angular velocity, which are given  by,
\begin{equation}\label{25}
M=\frac{a^{2}\sqrt{\pi}}{2\sqrt{S}}+\frac{\sqrt{\pi}Q^{2}}{2\sqrt{S}}+\frac{\sqrt{S}}{2\sqrt{\pi}}+
\frac{a^{2}\sqrt{S}}{2l^{2}\sqrt{\pi}}+\frac{S^\frac{3}{2}}{2l^{2}\pi^{\frac{3}{2}}}-
\frac{1}{2}\pi^{\frac{3\omega}{2}}S^{-\frac{3\omega}{2}}\alpha,
\end{equation}
and
\begin{equation}\label{26}
T=-\frac{a^{2}\sqrt{\pi}}{4S^{\frac{3}{2}}}-\frac{\sqrt{\pi}Q^{2}}{4S^{\frac{3}{2}}}+
\frac{1}{4\sqrt{\pi}\sqrt{S}}+\frac{a^{2}}{4l^{2}\sqrt{S}\sqrt{\pi}}+\frac{3\sqrt{S}}{4l^{2}\pi^{\frac{3}{2}}}+
\frac{3}{4}\pi^{\frac{3\omega}{2}}S^{-1-\frac{3\omega}{2}}\alpha\omega,
\end{equation}
While angular velocity remains unchanged and given by the equation (\ref{6}).\\
We will give a minor constant correction as $\epsilon$ to the corresponding action. So the corrected action will be as follows,
\begin{equation}\label{28}
\mathcal{I}=-\frac{1}{16\pi G}\int_{M}d^{4}x\{\sqrt{-g}(R-F^{2}+2(1+\epsilon)\times\Lambda)+L_{q}\}
\end{equation}
So,  we have modified thermodynamic quantities  for the corresponding  black hole in the following,
\begin{equation}\label{29}
M=\frac{a^{2}\sqrt{\pi}}{2\sqrt{S}}+\frac{\sqrt{\pi}Q^{2}}{2\sqrt{S}}+\frac{\sqrt{S}}{2\sqrt{\pi}}+
\frac{a^{2}\sqrt{S}}{2l^{2}\sqrt{\pi}}+\frac{S^\frac{3}{2}(1+\epsilon)}{2l^{2}\pi^{\frac{3}{2}}}-\frac{1}{2}
\pi^{\frac{3\omega}{2}}S^{-\frac{3\omega}{2}}\alpha,
\end{equation}
and
\begin{equation}\label{30}
T=-\frac{a^{2}\sqrt{\pi}}{4S^{\frac{3}{2}}}-\frac{\sqrt{\pi}Q^{2}}{4S^{\frac{3}{2}}}+\frac{1}{4\sqrt{\pi}\sqrt{S}}+
\frac{a^{2}}{4l^{2}\sqrt{S}\sqrt{\pi}}+\frac{3(1+\epsilon)\sqrt{S}}{4l^{2}\pi^{\frac{3}{2}}}+\frac{3}{4}
\pi^{\frac{3\omega}{2}}S^{-1-\frac{3\omega}{2}}\alpha\omega,
\end{equation}
angular velocity remains unchanged and given by the equation (\ref{6}).\\
Respecting the above thermodynamic expression, we will introduce some quantities  such as the conjugate to $\alpha$, which means the
\begin{equation}\label{eta}
\eta=-\frac{1}{2}\pi^{\frac{3\omega}{2}}S^{-\frac{3\omega}{2}},
\end{equation}
electric potential $\Phi=\frac{\sqrt{\pi}Q}{\sqrt{S}}$ and volume $V=\frac{4}{3}a^{2}\sqrt{\pi}S+\frac{4(1+\epsilon)S^{\frac{3}{2}}}{3\sqrt{\pi}}$ as well as $P=-\frac{\Lambda}{8\pi}$.
As before, the mass and the temperature of the black hole are modified with a small constant correction. It can be stated that when the correction parameter is continuously negative, the mass of the black hole decreases and the mass-charge ratio of the black hole decreases and approaches one. These changes give us a clue to the weak gravity conjecture, precisely like the previous section. Also, by using the equation (\ref{29}), the constant correction parameter $\epsilon$ is calculated as,
\begin{equation}\label{32}
\epsilon=-1-\frac{a^{2}l^{2}\pi^{2}}{S^{2}}-\frac{Q^{2}l^{2}\pi^{2}}{S^{2}}+\frac{2l^{2}
M\pi^{\frac{3}{2}}}{S^{\frac{3}{2}}}-\frac{a^{2}\pi}{S}-\frac{l^{2}\pi}{S}+l^{2}\pi^{\frac{3}{2}+
\frac{3\omega}{2}}S^{-\frac{3}{2}-\frac{3\omega}{2}}\alpha.
\end{equation}
Then, we take the derivative from $S$ so that we will have,
\begin{eqnarray}\label{33}
\frac{\partial\epsilon}{\partial S}&=&\frac{2a^{2}l^{2}\pi^{2}}{s^{3}}+\frac{2Q^{2}l^{2}\pi^{2}}{s^{3}}-\frac{3l^{2}M\pi^{\frac{3}{2}}}{S^{\frac{5}{2}}}\nonumber\\
&+&\frac{a^{2}\pi}{S^{2}}+\frac{l^{2}\pi}{S^{2}}+l^{2}\pi^{\frac{3}{2}+\frac{3\omega}{2}}S^{-\frac{5}{2}-\frac{3\omega}{2}}\alpha(-\frac{3}{2}-\frac{3\omega}{2})
\end{eqnarray}
Now, by combining the two equations (\ref{30}) and (\ref{33}), the first equation for universal relation is,
\begin{equation}\label{34}
-T\frac{\partial S}{\partial \epsilon}=\frac{S^{\frac{3}{2}}}{2l^{2}\pi^{\frac{3}{2}}}
\end{equation}
To obtain the second relation for completing this universal relation, we must use the $T=0$ and get the corresponding roots. The calculation for getting modified entropy is complicated, so we need to do some simplifications. Hence, we receive the extremal mass concerning $T=0$, so the equation (\ref{29}) leads us to arrive following equation,
\begin{equation}\label{35}
\frac{\partial M_{ext}}{\partial\epsilon}=\frac{S^{\frac{3}{2}}}{2l^{2}\pi^{\frac{3}{2}}}
\end{equation}
Here, we note that two equations (\ref{34}) and (\ref{35}) are the same. So, we first confirmed the Goon-Penco universal extremality relation for this black hole. To investigate another universal relation, just like the previous part concerning relation (\ref{32}), we have,
 \begin{equation}\label{36}
\frac{\partial \epsilon}{\partial Q}=-\frac{2l^{2}\pi^{2}Q^{2}}{S^{2}}
\end{equation}
By considering electric potential  $\Phi$, assuming extermality bound and using equation (\ref{36}), we will obtain the following equation,
\begin{equation}\label{37}
-\Phi\frac{\partial Q}{\partial \epsilon}=\frac{S^{\frac{3}{2}}}{2l^{2}\pi^{\frac{3}{2}}}
\end{equation}
We see that the equations (\ref{37}) and (\ref{35}) are the same. So, the second universal relation is also proven. In the following,  we seek to confirm other universal relations. So, respect to the pressure $P=\frac{3}{8\pi l^{2}}=-\frac{\Lambda}{8\pi}$ as well as the equation (\ref{32}) we obtain,
\begin{equation}\label{38}
\frac{\partial \epsilon}{\partial P}=\frac{3(a^{2}\pi+\pi Q^{2}-2M\sqrt{\pi}\sqrt{S}+S-\pi^{\frac{1}{2}+\frac{3\omega}{2}}\pi^{\frac{1}{2}-\frac{3\omega}{2}}\alpha)}{8P^{2}S^{2}}.
\end{equation}
According to the thermodynamic relation of  black hole as well as extremal bound, one can obtain,
\begin{equation}\label{39}
-V\frac{\partial P}{\partial \epsilon}=\frac{S^{\frac{3}{2}}}{2l^{2}\pi^{\frac{3}{2}}}
\end{equation}
Also, here we see that two equations (\ref{39}) and (\ref{35}) are the same. After confirming the universal relations stated in this section, given the rotation feature of the mentioned black hole, we also study the universal relationship associated with this feature of the black hole. So, according to the relation (\ref{32}) we have,
\begin{equation}\label{40}
\frac{\partial \epsilon}{\partial a}=-\frac{2a\pi(l^{2}\pi+S)}{S^{2}}.
\end{equation}
So, by using the equations (\ref{6}) and (\ref{40}), one can obtain,
\begin{equation}\label{41}
-\Omega\frac{\partial a}{\partial \epsilon}=\frac{S^{\frac{3}{2}}}{2l^{2}\pi^{\frac{3}{2}}},
\end{equation}
where $a$ can be understood as the black hole angular momentum per unit mass and $\Omega$ is its conjugate. As we can see,  two equations (\ref{41}) and (\ref{35}) are the same. After reviewing and confirming the universal relations for the charged-rotating-AdS black hole surrounded by quintessence, we want to examine the new universal relation related to the dark energy parameter. Hence, according to the equation (\ref{32}), we have the following equation,
\begin{equation}\label{42}
\frac{\partial \epsilon}{\partial\alpha}=l^{2}\pi^{\frac{3}{2}+\frac{3\omega}{2}}S^{-\frac{3}{2}-\frac{3\omega}{2}}
\end{equation}
So by using the $\eta$ and the equation (\ref{42}), we confirmed the last universal relation for this black hole, which is calculated as follow,
\begin{equation}\label{43}
-\eta\frac{\partial \alpha}{\partial\epsilon}=\frac{S^{\frac{3}{2}}}{2l^{2}\pi^{\frac{3}{2}}}=\frac{\partial M_{ext}}{\partial \epsilon},
\end{equation}
where $\eta$ is conjugate of $\alpha$. Here, also we draw plots and compare the two unmodified and modified masses (see Fig. \ref{2}). Therefore, we fix some parameters and plot the mass diagram in terms of $Q$ in Fig. \ref{2}. It can be seen that the mass ratio of the unmodified black hole to the amount of charge is more than one. We consider different modes of AdS space radius, fixed density of quintessence $\rho_{q}=1.0$ and $\omega=-\frac{2}{3}$. As we can see in Fig. \ref{2} (a), the mass curve is different from the charge of the black hole for different modes. We will perform a numerical analysis of the black hole's mass and describe the entropy. As we can see in Fig. \ref{2} (b), we compare that with its unmodified state by using a small constant correction for the mass. As shown in Fig. \ref{2} (b),  when we consider the small correction as negative, the mass of a black hole decreases to one, and in the same way, the charge-to-mass ratio increases, which is satisfied by the weak gravity conjecture. As we can see, the obtained results show us the quintessence terms play a very effective role in calculating the mass-charge ratio and concept of weak gravity conjecture of black holes.

\begin{figure}[h!]
 \begin{center}
 \subfigure[]{
 \includegraphics[height=6cm,width=6cm]{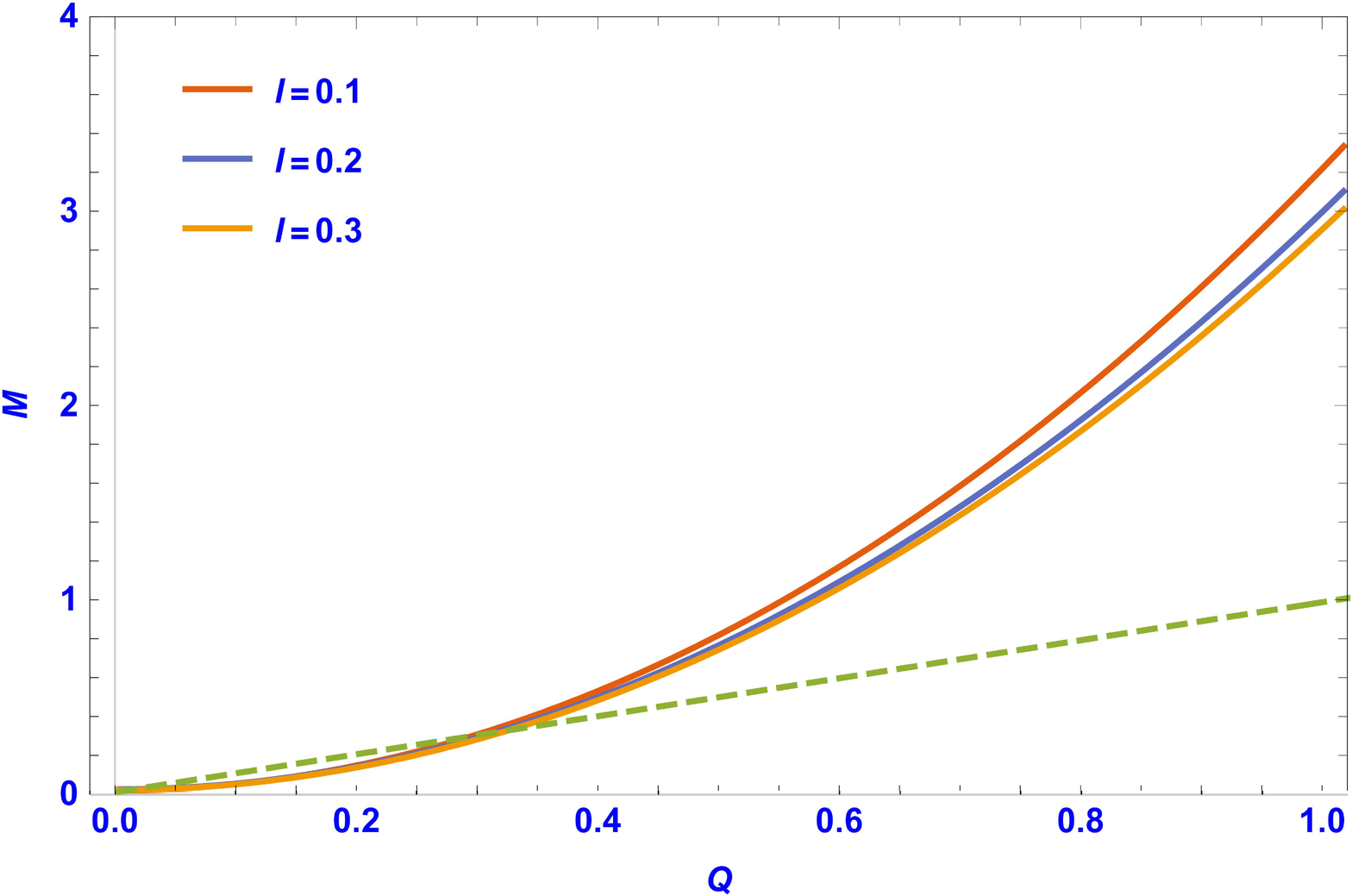}
 \label{2a}}
 \subfigure[]{
 \includegraphics[height=6cm,width=6cm]{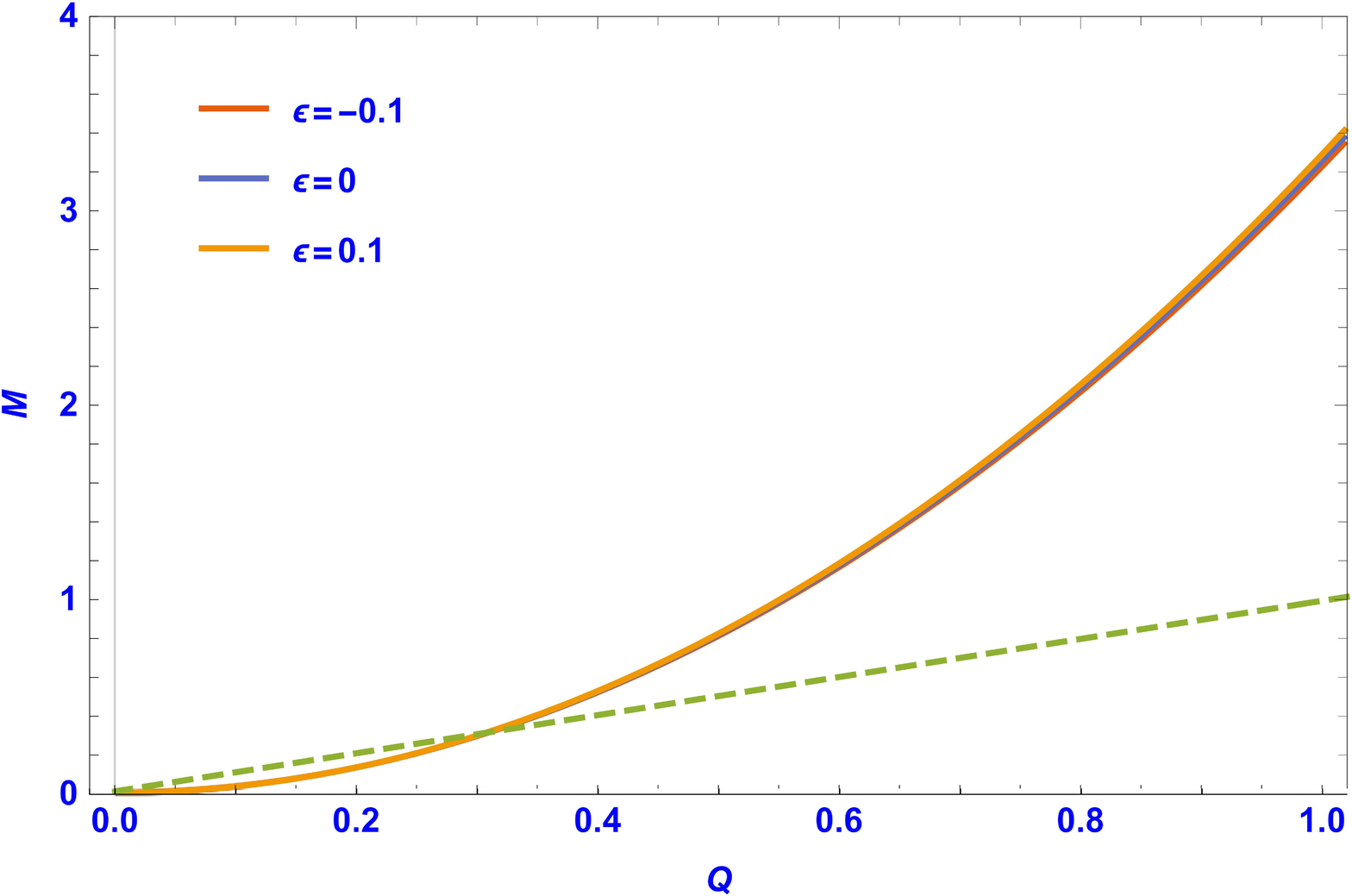}
 \label{2b}}
  \caption{\small{The plot of $M$ in terms of $Q$ with $\omega=-\frac{2}{3}$, for (a) $\epsilon=0$ (unmodified mass); (b) $l=0.1$ (modified mass). Dashed lines represent extremal case of Kerr-Newman-AdS black hole with quintessence.}}
 \label{2}
 \end{center}
 \end{figure}

\section{Kerr-Newman-AdS black hole with quintessence and cloud of strings}
In this section, we will do the same as the previous two sections, except that in this section, we will add another term, i.e., the cloud of strings, to the Kerr-Newman-AdS black hole surrounded by quintessence. Therefore, the metric for Kerr-Newman-AdS black hole with quintessence and cloud of strings is expressed in the following form \cite{54},
\begin{eqnarray}\label{44}
ds^{2}&=&\frac{\Sigma^{2}}{f(r)}dr^{2}+\frac{\Sigma^{2}}{g(\theta)}d\theta^{2}
+\frac{g(\theta)\sin^{2}\theta}{\Sigma^{2}}\left(a\frac{dt}{\Xi}-(r^{2}+a^{2})\frac{d\phi}{\Xi}\right)^{2}\nonumber\\
&-&\frac{f(r)}{\Sigma^{2}}\left(\frac{dt}{\Xi}-a\sin^{2}\theta\frac{d\phi}{\Xi}\right)^{2}
\end{eqnarray}
where
\begin{eqnarray}\label{45}
f(r)&=&(1-b)r^{2}+a^{2}+Q^{2}-2Mr-\frac{\Lambda}{3}r^{2}(r^{2}+a^{2})-\alpha r^{1-3\omega_{q}}\nonumber\\
g(\theta)&=&1+\frac{\Lambda}{3}a^{2}\cos^{2}\theta,
\end{eqnarray}
with the same definition as (\ref{24-2}).\\
Using the  $f(r)=0$, we obtain the temperature, entropy, etc., in terms of horizon radius. We apply a small constant correction to the action and calculate the modified thermodynamic relation for the corresponding black hole. So, the same as before, we have the following equations,
\begin{equation}\label{46}
M=\frac{a^{2}\sqrt{\pi}}{2\sqrt{S}}+\frac{\sqrt{\pi}Q^{2}}{2\sqrt{S}}+\frac{\sqrt{S}}
{2\sqrt{\pi}}-\frac{b\sqrt{S}}{2\sqrt{\pi}}+\frac{a^{2}\sqrt{S}}{2l^{2}\sqrt{\pi}}+
\frac{(1+\epsilon)S^{\frac{3}{2}}}{2l^{2}\pi^{\frac{3}{2}}}-\frac{1}{2}
\pi^{\frac{3\omega}{2}}S^{-\frac{3\omega}{2}}\alpha,
\end{equation}
and
\begin{equation}\label{47}
T=-\frac{a^{2}\sqrt{\pi}}{4S^{\frac{3}{2}}}-\frac{\sqrt{\pi}Q^{2}}{4S^{\frac{3}{2}}}+
\frac{1}{4\sqrt{S\pi}}-\frac{b}{4\sqrt{S\pi}}+\frac{a^{2}}{4l^{2}
\sqrt{S\pi}}+\frac{3(1+\epsilon)\sqrt{S}}{4l^{2}\pi^{\frac{3}{2}}}+
\frac{4}{3}\pi^{\frac{3\omega}{2}}S^{-1-\frac{3\omega}{2}}\alpha \omega.
\end{equation}
Then, we will investigate some quantities with respect to thermodynamic relation such as,
\begin{equation}\label{eta2}
\eta=-\frac{1}{2}\pi^{\frac{3\omega}{2}}S^{-\frac{3\omega}{2}}
\end{equation}
together electric potential $\Phi=\frac{\sqrt{\pi}Q}{\sqrt{S}}$ and black hole volume $V=\frac{4\sqrt{S}(a^{2}\pi+S+S\epsilon)}{3\sqrt{\pi}}$ as well as $P=-\frac{\Lambda}{8\pi}$ and the important relation for this section
\begin{equation}\label{zeta}
\zeta=-\frac{\sqrt{S}}{2\sqrt{\pi}}
\end{equation}
which is conjugate to $b$. We find that when the added correction is continuously negative, the mass of the black hole decreases, and the mass-charge ratio of the black hole decreases and approaches one. These changes give us a clue to the weak gravity conjecture. By solving the equation (\ref{46}), the constant correction parameter $\epsilon$ is given by the following equation,
\begin{equation}\label{49}
\epsilon=-1+\frac{-a^{2}\pi(^{2}\pi+S)+l^{2}\pi(-\pi Q^{2}+2M\sqrt{\pi S}+(-1+b)S+\pi^{\frac{1}{2}+\frac{3\omega}{2}}S^{\frac{1}{2}-\frac{3\omega}{2}}\alpha)}{S^{2}}.
\end{equation}
So, we take the derivative with respect to $S$, and obtain,
\begin{equation}\label{50}
\frac{\partial\epsilon}{\partial S}=\frac{2a^{2}l^{2}\pi^{2}+2l^{2}\pi^{2}Q^{2}-3l^{2}M\pi^{\frac{3}{2}}\sqrt{S}+a^{2}\pi S+l^{2}\pi S-bl^{2}\pi S-\frac{3}{2}l^{2}\pi^{\frac{3}{2}+\frac{3\omega}{2}}S^{\frac{1}{2}-\frac{3\omega}{2}}\alpha(1+\omega)}{S^{3}}.
\end{equation}
Now, by combining the equations (\ref{47}) and (\ref{50}), one can obtain the following equation,
\begin{equation}\label{51}
-T\frac{\partial S}{\partial \epsilon}=\frac{S^{\frac{3}{2}}}{2l^{2}\pi^{\frac{3}{2}}}
\end{equation}
To obtain the second relation for completing the corresponding universal relation, we assume $T=0$ and use the temperature equation. But this calculation is complicated and needs to be simplified the obtain modified entropy. Hence,  we employ  extremal mass and use the equation (\ref{47}) to get,
\begin{equation}\label{52}
\frac{\partial M_{ext}}{\partial\epsilon}=\frac{S^{\frac{3}{2}}}{2l^{2}\pi{\frac{3}{2}}}.
\end{equation}
Two equations (\ref{51}) and (\ref{52}) are precisely the same. We first confirm the Goon-Penco universal extremality relation for this black hole. Now, we want to confirm the other universal relation, likewise the previous sections. So concerning relation (\ref{49}), we will have,
\begin{equation}\label{53}
\frac{\partial \epsilon}{\partial Q}=-\frac{2l^{2}\pi^{2}Q}{S^{2}}.
\end{equation}
Also, by using the equation (\ref{53}), the electric potential $\Phi$ and extremality bound, we have,
\begin{equation}\label{54}
-\Phi\frac{\partial Q}{\partial \epsilon}=\frac{S^{\frac{3}{2}}}{2l^{2}\pi{\frac{3}{2}}}.
\end{equation}
As we can see the equation (\ref{54}) and (\ref{52}) are same.  In the following, we use the pressure $P=\frac{3}{8\pi l^{2}}=-\frac{\Lambda}{8\pi}$ and the equation (\ref{49}), one can obtain,
\begin{equation}\label{55}
\frac{\partial P}{\partial \epsilon}=-\frac{3S}{8l^{2}\pi(a^{2}\pi+S+S\epsilon)}.
\end{equation}
Here, according to the thermodynamic relation of  black hole such as $V$ as well as extremal bound one can obtain following equation,
\begin{equation}\label{56}
-V\frac{\partial P}{\partial \epsilon}=\frac{S^{\frac{3}{2}}}{2l^{2}\pi{\frac{3}{2}}}.
\end{equation}
It is clear that the equations (\ref{56}) and (\ref{52}) are exactly the same. Also, according to relation (\ref{49}), we have
\begin{equation}\label{57}
\frac{\partial \epsilon}{\partial a}=-\frac{2a\pi(l^{2}\pi+S)}{S^{2}}
\end{equation}
So by using the equations (\ref{6}) and (\ref{57}), one can obtain,
\begin{equation}\label{58}
-\Omega\frac{\partial a}{\partial \epsilon}=\frac{S^{\frac{3}{2}}}{2l^{2}\pi{\frac{3}{2}}}.
\end{equation}
As we see here, the two equations (\ref{58}) and (\ref{52}) are the same. Now, we are going to examine another universal relation that is related to the dark energy parameter. Hence, according to the equation (\ref{49}), one can have,
\begin{equation}\label{59}
\frac{\partial \epsilon}{\partial\alpha}=l^{2}\pi^{\frac{3}{2}(1+\omega)}S^{-\frac{3}{2}(1+\omega)}.
\end{equation}
So, by using the parameter $\eta$ and the equation (\ref{59}), we confirmed the other relation for this black hole which is calculated by,
\begin{equation}\label{60}
-\eta\frac{\partial \alpha}{\partial\epsilon}=\frac{S^{\frac{3}{2}}}{2l^{2}\pi{\frac{3}{2}}}.
\end{equation}
As we can see,  the equations (\ref{60}) and (\ref{52}) are extremely the same. After proving all the universal relations, we now obtain a new universal relation from the black hole's new property as the cloud of string. We prove the corresponding universal relation and show how such a black hole is satisfied by weak gravity conjecture. So, we use the relation (\ref{49}) and obtain,
\begin{equation}\label{61}
\frac{\partial \epsilon}{\partial b}=\frac{l^{2}\pi}{S}.
\end{equation}
Here, the above equation and relation of $\zeta$ in-text lead us to confirm the last universal relation. So, it is given by,
\begin{equation}\label{62}
-\zeta\frac{\partial b}{\partial \epsilon}=\frac{S^{\frac{3}{2}}}{2l^{2}\pi{\frac{3}{2}}}=\frac{\partial M_{ext}}{\partial\epsilon}.
\end{equation}

\begin{figure}[h!]
 \begin{center}
 \subfigure[]{
 \includegraphics[height=6cm,width=6cm]{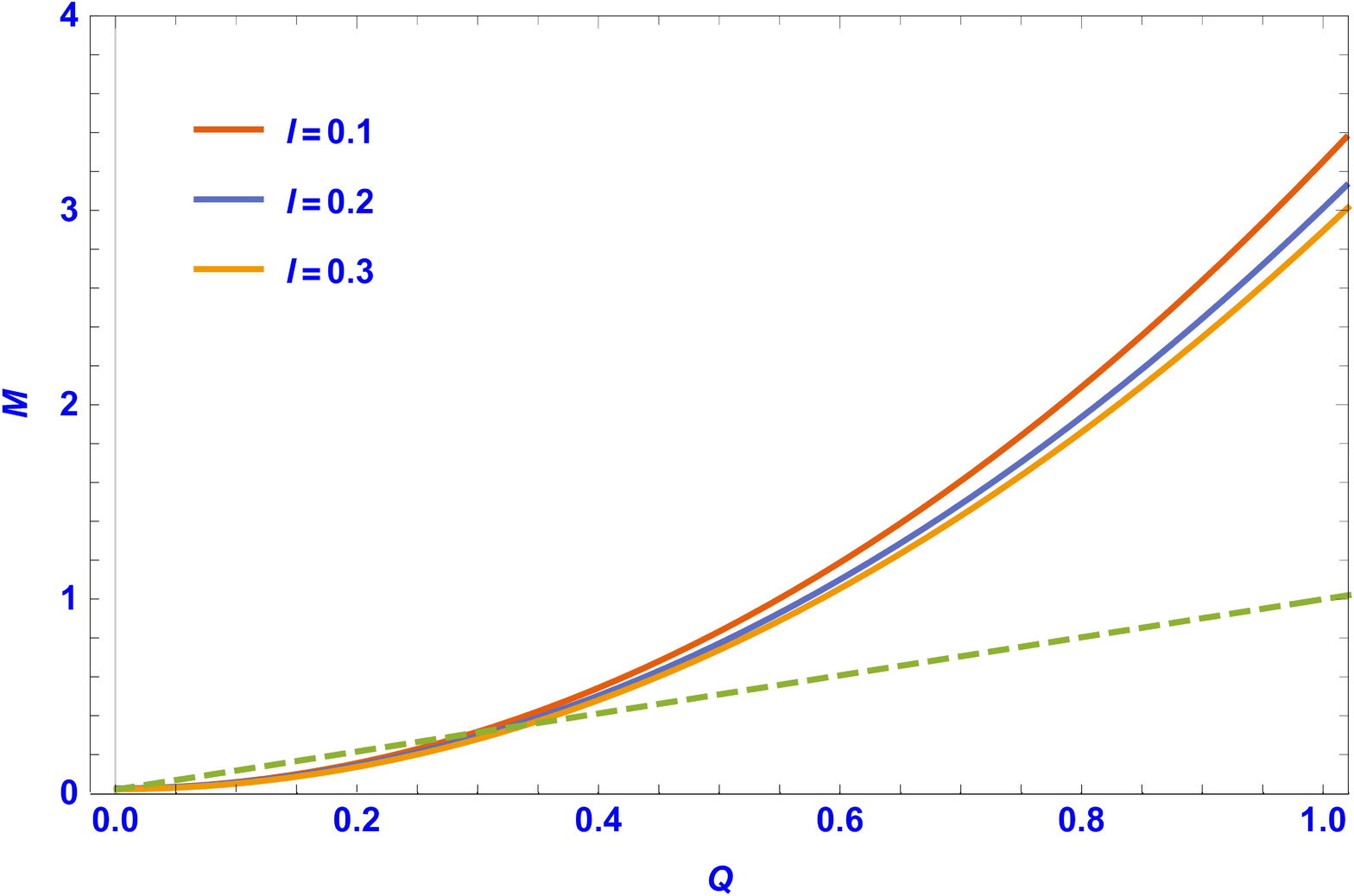}
 \label{3a}}
 \subfigure[]{
 \includegraphics[height=6cm,width=6cm]{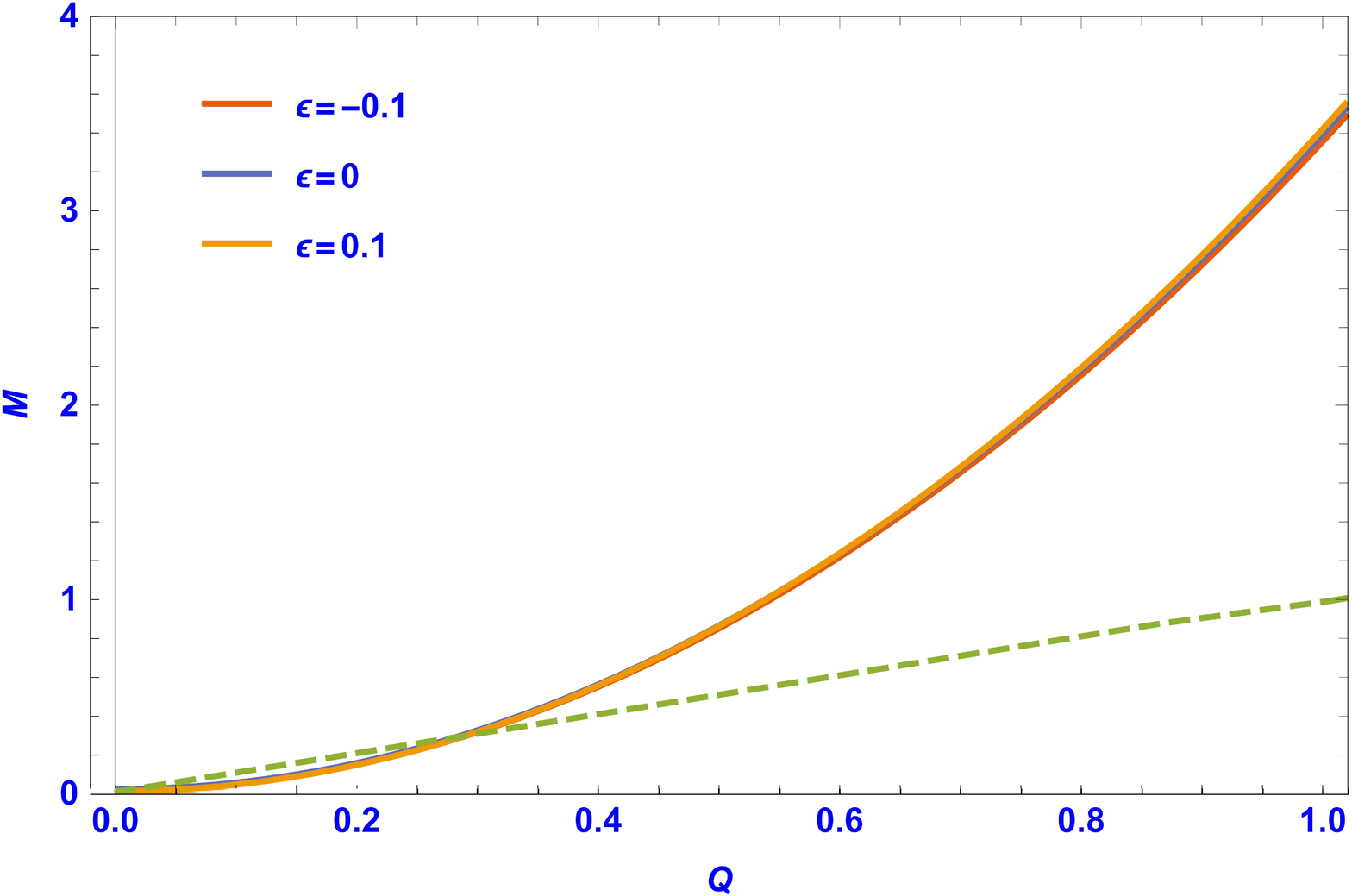}
 \label{3b}}
  \caption{\small{The plot of $M$ in terms of $Q$ with $\omega=-\frac{2}{3}$, for (a) $\epsilon=0$ (unmodified mass); (b) $l=0.1$ (modified mass). Dashed lines represent extremal case of Kerr-Newman-AdS black hole with quintessence and cloud of strings.}}
 \label{3}
 \end{center}
 \end{figure}

Here the same as before, we have plots of Fig. \ref{3}. These show us how the cloud of string terms will affect the corresponding system. Therefore, we can see from all the above relations that we examined the universal relationships for Kerr-Newman-AdS black holes in the three forms. We obtained the universal relations in the different cases and compared them with each other. Using thermodynamic relations,  we observe that when a new property is added to the black hole, a new universal relation can be considered. In this paper, we also calculated the new universal relations by assuming the rotating and quintessence dark energy and the cloud of the string of the black hole. We observed that this universal relation is well established. As it turned out when the constant correction is negative, the mass decreases, and the mass-charge ratio also decreases to equal one. There is a sign of weak gravity conjecture behavior, which was examined for all cases concerning the extremal bound. Therefore, by considering new modes such as high dimensions and black holes with different structures, new universal relations can be evaluated, and we will study this problem in the future.

\section{Conclusion}
The general relativity corrections lead to a glancing relationship between entropy and extremality bound. This relationship has been investigated for some kinds of black holes, such as charged AdS, rotating, and massive gravity black holes. This article confirmed these universal relations for a charged-rotating-AdS black hole, which adds a small constant correction to the action. Then, we examined these calculations for this black hole while surrounded by the quintessence and quintessence with the cloud of strings. We also evaluated a new universal relation between the mass of the black hole and the factor related to quintessence density. Also, we got a new universal relation for the cloud of string, which is given by,
\begin{equation*}
-\zeta\frac{\partial b}{\partial \epsilon}= \frac{\partial M_{ext}}{\partial\epsilon}
\end{equation*}
We observed that this universal relation is well established. We know that the added constant correction is inversely related to the entropy of the black hole. In that case, the mass-to-charge ratio decreases. This leads us to conclude that the black hole has a WGC-like behavior. Here we note that calculating universal relations by considering some features such as higher dimensions, strings fluid mimics, van der Waals fluid behavior, and black holes with other structures such as Einstein-Gauss-Bonnet may be exciting research for the future.\\\\

\end{document}